# A DEEP MULTICOLOR SURVEY II. INITIAL SPECTROSCOPY AND COMPARISON WITH EXPECTED QUASAR NUMBER COUNTS

Patrick B. Hall[1]

Steward Observatory, University of Arizona, Tucson AZ 85721

E-mail: phall@as.arizona.edu

Patrick S. Osmer

Astronomy Department, The Ohio State University, 174 West 18th Avenue, Columbus, OH 43210

E-mail: posmer@payne.mps.ohio-state.edu

Richard F. Green, Alain C. Porter[2]

National Optical Astronomy Observatories, P.O. Box 26732, Tucson AZ 85726

E-mail: rgreen@noao.edu

Stephen J. Warren

Imperial College, Astrophysics Group, The Blackett Laboratory, Prince Consort Road, London SW7 2BZ

E-mail: s.j.warren@ic.ac.uk

*Subject headings:* Surveys, Quasars: General, Galaxies: Compact

## ABSTRACT

We have used the KPNO 4-meter Mayall telescope to image 0.83 square degrees of sky in six fields at high galactic latitude in six filters spanning 3000-10000Å to magnitude limits ranging from 22.1 to 23.8. As a first use of this database, we have conducted a multicolor survey for quasars. We discuss various methods of selecting outliers in different color-color diagrams and multicolor space that have been used to identify quasars at all redshifts from their colors alone. We discuss the initial results of our program of spectroscopic identification which has so far resulted in the identification of over forty faint quasars, including one at z>4, a similar number of compact narrow-emission-line galaxies, and a number of unusual and potentially interesting stars. We use these spectroscopic results, along with extensive simulations

---

of quasar spectra, to study the efficiency of our candidate selection procedures. Finally, we compare the number counts of our quasars and quasar candidates to the expected numbers based on previous studies of the quasar luminosity function. The agreement of our observations with these expectations is good in most cases. However, we do estimate that our survey contains more quasars with B<21 and z<2.3 than expected from the results of Koo & Kron (1988) and more z>3 quasars than expected from the results of Warren, Hewett, & Osmer (1994), both at the $3\sigma$ level. Additional spectroscopic observations will be required to confirm or refute these excesses.

## 1. Introduction

This is the second in a series of papers presenting results from a deep multicolor optical survey carried out at the KPNO 4-meter telescope. The survey was conducted in six filters spanning 3000-9000Å in six different high-Galactic-latitude fields. The primary objective was to construct a well-defined stellar locus from which quasar candidates could be easily isolated via their unusual colors. However, many other interesting questions can be addressed with the survey. We achieved good photometric calibration, and the star-galaxy classification effects are well understood, thus the survey should be useful for studies of Galactic star counts (Reid & Majewski 1993). The large number of galaxies detected in the survey also forms a valuable database for the study of field and emission-line galaxies. Our wide wavelength coverage makes the survey especially useful for studies of field galaxy color evolution and for the identification of rare galaxy types such as E+A galaxies (Liu et al. 1996).

The survey for quasars has two principal objectives: 1) to investigate the luminosity function of high-redshift quasars (z>3) at fainter magnitudes than has been done before, for which the absolute magnitudes were more comparable with previous surveys at lower redshift; and 2) to determine the luminosity function of quasars with $z \leq 2.2$ at faint enough magnitudes to constrain the debate on luminosity vs. density evolution, which has been unresolved for lack of data at the faint end of the luminosity function, where the differences between the two hypotheses are pronounced.

The assembly and photometric calibration of the catalog of stellar objects is detailed in Paper I (Hall et al. 1996). The catalog contains 21,375 stellar objects from 0.83□° of sky, with magnitudes in six filters: standard UBV, a narrow R filter that, apart from a zeropoint shift, is essentially equivalent to a standard Harris R, and two special I filters (I75 and I86) which each cover about half the wavelength range of the standard I filter (see Paper I for details). The stellar object catalog is 90% complete and 20-33% contaminated by misclassified galaxies at the $5\sigma$ detection limits, depending on the filter (see Table 3 of Paper I). From this catalog we wish to isolate as many as possible of the ∼200 quasars with B≤22.6 expected in our survey area (Majewski et al. 1993), with a minimum of contamination from stars and other objects. Careful candidate selection is obviously necessary, and in this paper we discuss our methods and preliminary spectroscopic results. Paper III (Osmer et al. 1996) will present final spectroscopic results and a full derivation of the luminosity function.

This paper is organized as follows. In §2 we describe the general principles behind the different multicolor selection criteria we used. We describe the spectroscopic observations we made with the HYDRA multifiber spectrograph on the KPNO 4-m telescope in §3 and present the spectroscopic results in §4. We describe in §5 the extensive simulations of quasar spectra we carried out and used to refine our selection criteria (§6) and calculate their detection probabilities (§7). In §8 we use these detection probabilities and the spectroscopic results to compare the numbers of candidates in our survey with the values expected from other surveys. Finally, §9 contains a summary of our results and conclusions.

## 2. Multicolor Candidate Selection

### 2.1. General Overview

Multicolor selection is one of the oldest methods of separating candidate quasars from stars in an observational database. Essentially, the technique relies on the fact that the spectral energy distributions (SEDs) of quasars are, on the average, very different from the SEDs of stars. Broadband photometry can distinguish adequately between the two sorts of objects, and stars and quasars at many redshifts tend to be rather well separated when plotted in color-color diagrams.

An example of a useful color-color diagram is the U-B versus B-V diagram, plotted in Figure 1 for two magnitude bins. We refer to color-color diagrams with the notation U-B/B-V, indicating the colors plotted on the y and x axes respectively. We plot color-color diagrams with the bluer color on the y axis, so that the bluest objects appear in the upper left-hand corner, and the reddest in the lower right-hand corner. The observed stellar locus runs from F to M stars from the upper left to lower right, with the rare (at high galactic latitude) stars earlier than F extending even farther to the upper left. In all our color-color diagrams, objects from the catalog are plotted as error bars, with only half the full error bars plotted, for clarity. Error bars extending off the diagram indicate upper limits. Simulated quasar colors (discussed in §5) are plotted as dots, spectroscopically confirmed quasars as solid boxes, spectroscopically confirmed galaxies as open boxes, spectroscopically confirmed stars as inverted triangles, and observed but unidentified candidates as crosses. The colors of two simulated quasars are plotted as a function of redshift in both plots, with a few redshifts marked. In Figure 1a, the solid line is a quasar with spectral index $\alpha=-0.25$ and strong emission lines, and the dashed line is a quasar with spectral index $\alpha=-0.75$ and average emission lines. In Figure 1b, the solid line is a quasar with spectral index $\alpha=-0.25$ and weak emission lines, and the dotted line is a quasar with spectral index $\alpha=-1.25$ and strong emission lines. These tracks span the range of typical emission-line strengths and spectral indices for quasars.

It is seen in the U-B/B-V diagram of Figure 1a that most quasars are found at z<3 and appear bluer than stars. This explains why the oldest color selection method for isolating quasars is the UV-excess (UVX) criterion of U-B≤-0.3. This simple but effective criterion for selecting quasars with redshifts ranging up to 2.8 does suffer some contamination from objects such as O and B stars, white dwarfs, hot metal-poor stars (e.g. halo subdwarfs; class sdO or sdB), and compact narrow emission-line galaxies (CNELGs). However, CNELGs typically are redder in B-V than quasars due either to the presence of lines such as H$\beta$, H$\gamma$, and [OII] 3727 in the V filter or to the contribution of the stellar population from the galaxy, and this helps to discriminate them from quasars.

In Figure 1b, at magnitudes where photometric errors begin to increase, the simple criterion of U-B<-0.3 begins to be contaminated by F stars with intrinsic colors of U-B∼-0.15 and B-V∼0.5 that are scattered by errors to different values of U-B and B-V. At the red end of the stellar locus, the figure shows how magnitude limits begin to affect the areas from which we can select outliers. Similar effects happen in all color-color diagrams at magnitudes where photometric errors are no longer negligible. This makes it necessary to devise selection criteria that are functions of magnitude, which we have done by binning the objects in magnitude.

The simple UVX criterion U-B<-0.3 breaks down rapidly for quasar redshifts z>2.2, when the U-B and then B-V colors redden as Ly-$\alpha$ redshifts into the B and then V bands and the Ly-$\alpha$ forest begins to weaken the flux in U and then B. However, a succession of different color-selection criteria can be constructed in order to take advantage of the observed position of Ly-$\alpha$ and the absorption blueward of it at various redshifts. For example, Irwin et al. (1991) devised a very successful criterion for the selection of 4.0<z<4.5 quasars by searching for a very red (B-R>3) color caused by the presence of Ly-$\alpha$ emission in the R band and the substantial depression of the quasar continuum in the B band due to intervening absorption. Such quasars will be bluer in R-I than late-type stars due to the presence of Ly-$\alpha$ emission in R. Thus objects that are very red in B-R and somewhat blue in R-I are strong z>4 quasar candidates. This 'BRX' criterion can be roughly generalized to other redshift ranges and filters as well. Quasars with 3.1<z<3.9 will appear red in U-V or B-V and blue in V-R, since Ly$\alpha$ is in the V band. Quasars with 4.8<z<5.6 will have Ly$\alpha$ in the I75 band and will be red in V-I75 and blue in I75-I86, while late-type stars will generally be red in both V-I75 and I75-I86.

A more general method of selecting quasar candidates from a multicolor data set is to make use of all six independent magnitudes (or five independent colors) simultaneously, instead of using only two colors at a time in a color-color diagram. This can be done considering the positions of the objects in a six-dimensional magnitude space (or five-dimensional color space), as in Warren et al. 1991. The basic criterion used in selecting outliers in multidimensional spaces is the $n$th-nearest-neighbor distance. Every object in a catalog is compared to every other object, and the distance between objects i and j can be calculated in one of several ways. For example, in a six-dimensional magnitude space where the objects' positions are defined by their UBVRI75 and

I86 magnitudes, the *signal distance* $d_S$ is given by:

$$d_{Sij}^2 = U_{ij}^2 + B_{ij}^2 + V_{ij}^2 + R_{ij}^2 + I75_{ij}^2 + I86_{ij}^2 \tag{1}$$

whereas the *signal-to-noise distance* $d_{SN}$ is given by:

$$d_{SNij}^2 = (\frac{U_{ij}}{\sigma_{Uij}})^2 + (\frac{B_{ij}}{\sigma_{Bij}})^2 + (\frac{V_{ij}}{\sigma_{Vij}})^2 + (\frac{R_{ij}}{\sigma_{Rij}})^2 + (\frac{I75_{ij}}{\sigma_{I75ij}})^2 + (\frac{I86_{ij}}{\sigma_{I86ij}})^2 \tag{2}$$

For each object i, the $n$th smallest distance (of either $d_S$, $d_{SN}$, or some combination thereof) is chosen to be a measure of its 'outlyingness', where $n$ is usually chosen to be a number such as 10, in order to reduce the possibility of missing small groups of outliers.

However, selection of outliers in six-dimensional magnitude space is difficult for a survey such as ours which does not contain a large number of bright objects. This is because bright stars of even slightly unusual color will be outliers in magnitude space when there are only a few such bright stars in the catalog. Selection in color-color space is better, but at faint magnitudes, larger errors lead to considerably more scatter in color-color space than at bright magnitudes and it becomes easier for stars to contaminate the catalog of outliers.

One obvious solution is to use magnitude-dependent outlier criteria. This is exactly analogous to what we have done with simple color-color diagrams by binning the sample by magnitude and using different color-color selection criterion for each magnitude bin. The analogous procedure in continuous (as opposed to binned) color-color-magnitude space was developed by S. Warren.

The essence of the procedure is to measure the 10th-nearest-neighbor distances $d_S$ and $d_{SN}$ for each object in the color space defined by (U-B,B-V,V-R,R-I75,I75-I86) within a magnitude range m-2<m<m+0.2, where m is the object's I75 apparent magnitude. The quantity $d_S \times d_{SN}$ was used to rank each object's outlyingness, on the assumption that slightly unusual bright objects with small errors, and thus large $d_{SN}$, will have smaller $d_S$ than true outliers, and that faint outliers with large errors, and thus small $d_{SN}$, will have larger $d_S$. This procedure has not been optimized, but worked quite well; the multicolor-space candidates included ∼90% of the spectroscopically identified quasars, but only ∼25% of the spectroscopically identified compact galaxies and only ∼60% of the spectroscopically identified stars. Thus a rough estimate is that the current unoptimized multicolor-space technique is about ∼90% as complete as the (also unoptimized) color-color diagram selection technique, but is ∼50% more efficient; i.e. in a given number of candidates it contains ∼50% more quasars. It will not be discussed further in this paper, but it is clearly an important and efficient technique for identification of candidates which we will continue to improve upon and use.

## 2.2. Initial Candidate Selection

Our primary targets were selected using the multicolor-space outlier selection algorithm and selection of outliers from a number of color-color diagrams: U-B/B-V, U-V/V-R and B-V/V-R,

B-R/R-I75 and B-R/I75-I86, and several diagrams involving V,R,I75 and I86, intended to find rare but valuable $z>5$ objects. As we found no $z>5$ objects, we will not discuss the selection of such candidates here. (Although the BRX technique is theoretically efficient at selecting quasars out to $z>5$, the R-band k-correction rapidly becomes large at such redshifts, and the luminosity required for such a quasar to still be detected in R would be implausible even for a survey over a much larger area than ours.)

It is possible to search all outlying regions of multicolor space for quasars, but the SEDs of quasars, though they do show tremendous variety, are such that not all regions of multicolor space are equally likely to contain quasars. Thus, to improve the efficiency of our quasar candidate selection, we generated synthetic quasar spectra, computed their colors in our filter set, and used them to identify the best areas on the various color-color diagrams from which to select candidates for spectroscopic followup. (These simulations are discussed at length in §5, and the refined selection criteria in §6.) As a lower priority we also explored other outlying regions of color-color space and objects with only one unusual color (i.e. objects located in the stellar locus in most color-color diagrams) to help ensure that unusual types of quasars were not missed. Neither our survey nor that of Warren et al. (1991) found any unusual population of quasars that did not reside in expected areas of color-color space. Of course, this only sets an upper limit on the surface density of unusual quasars at these magnitudes. Substantial unusual populations of quasars may still exist, if their properties are such that they are difficult to find in optical surveys (e.g. the dust-reddened quasars of Webster et al. (1995)).

Objects selected as candidates were visually inspected in order to ensure that their colors were not corrupted by some defect in the images. All images relevant to how the candidate was selected were inspected. For example, UVX-selected objects were inspected on the U, B, and V images of the field. Inspection of other images is irrelevant because, for example, a bad pixel in R will not affect the U-B color of an object except through a neglible color-term effect. Spurious candidates are typically caused by centering errors in the photometry or by coincidence of an object with a cosmic ray, bad column, or the trail of a satellite or asteroid. Note that objects with these various problems will tend to be overrepresented in candidate lists because an erroneous measurement is much more likely to scatter an object into an outlying area of color-color space than into another part of the stellar locus, simply because the fraction of color-color space inhabited by stars is relatively small.

## 3. Spectroscopy

We used the HYDRA multifiber spectrograph on the KPNO 4-meter Mayall telescope to obtain spectroscopy of candidate quasars. HYDRA is an automated fiber positioner that allows placement of up to ∼97 fibers in a ∼50′ field of view. (see Barden et al. (1993) for a detailed description). Optimal fiber assignments were generated using software developed by Massey (1992). We had successful observing runs with HYDRA in 1992 Oct 1-3 and 1993 May 27-29. We used the blue fiber cable with the KPC-10 grating (316 lines/mm) to obtain spectral coverage from 3500 to 7500 Å at a resolution of 6-8 Å (∼2 Å/pixel). We used the Tektronix T2Kb 2048$x$2048 CCD (24$\mu$ pixels) and a gain of 4.2 e-/ADU. On the last night of the May 1993 run we used the red fiber cable with the BL-181 grating (316 lines/mm) with spectral coverage from 5500 to 9500 Å (∼2 Å/pixel) at a resolution of ∼8 Å.

Reduction of the HYDRA data was performed by P. Osmer and A. Porter using the DOHYDRA tasks (Valdes 1992) as implemented in IRAF V2.10.3. The reduction steps taken were as follows:

- the CCD images containing the raw two-dimensional spectra were processed in the usual manner to make overscan, bias, and dark current corrections.

- the images were cleaned of cosmic rays by two different algorithms in order to check later on the effectiveness of the cosmic ray removal and the reality of putative emission line detections.

- the scattered light was modeled by a high-order two-dimensional polynomial surface and subtracted from the images.

- a flatfield spectrum was constructed using domeflat observations with the same fiber setup used on the sky, with the fiber-to-fiber throughput variations normalized by twilight sky observations.

- the spectra were optimally extracted, collapsed to a one-dimensional spectrum, flatfielded, and wavelength-calibrated using observations of a comparison lamp, again in the same setup as used on the sky.

- a mean sky spectrum was interactively constructed from the individual sky fiber spectra, rejecting obvious outliers either with flux contaminated by an object in or near the random fiber position or with vignetting by the guide probe or by being located too close to the edge of the field of view.

- the mean sky spectrum was then subtracted from all the one-dimensional spectra.

The spectra were inspected independently by P. Osmer and P. Hall and classified (either definitely or tentatively) as star, galaxy, or quasar. Redshifts or rough stellar classifications, as appropriate, were estimated based on emission and/or absorption features, and the redshifts themselves were also classified as definite or tentative. The agreement in object detection and classication and redshift calculation between the two independent examinations of the spectra was quite good, and agreement was always reached on the questionable objects after further collaborative review.

## 4. Initial Spectroscopic Results

With all fields except the 10e field observed to date, and with typically $\gtrsim 4$ hours of total integration time on each object, we have spectroscopically identified objects as faint as B=22.3, V=22.0, and R=21.6. Among our candidates we have identified 46 quasars, 39 galaxies, and ~100 stars. The other ~150 of the ~350 candidates observed spectroscopically to date are unidentified because of low S/N.

### 4.1. Quasars

Table 1 lists positions, redshifts, magnitudes, and notes on the quasars and quasar candidates. Object designations are given as DMS (Deep Multicolor Survey) followed by the epoch 1950.0 position. The redshift distribution of the 44 quasars with tentative or secure redshifts is shown in Figure 2. The redshifts span the range 0.68<z<4.33. There are 8 objects with 2<z<3; 4 objects with z>3, and one at z=4.33.

The minimum redshift of 0.68 is likely a consequence of the faint magnitude limit of the survey. Otherwise we would expect to see quasars at z=0.3, where the MgII 2798 line is in the U band, and where there is a strong peak in the redshift distribution for the Hewitt & Burbidge (1993) catalog. However, at z=0.3 and a B magnitude of 21, $M_B = -19.46$ ($H_0 = 75$, $q_0 = 0.5$) and the host galaxy is bright enough for the object to be classified as a galaxy, not a star-like object.

Several quasars have tentative redshifts, based on only one definite line or on low S/N detections of only one or two lines; all such objects have colors consistent with their tentative redshifts. Two objects show tentative broad line(s) that cannot be assigned a definite redshift: DMS1358-0100 has colors consistent with a z<2.8 quasar while DMS1358-0045 could be a quasar or hot star based on its colors. Several other objects with featureless continua have also eluded classification; most are likely to be stars, but some may be weak-lined quasars or BL Lacs.

Figure 3 displays the spectra of the two quasars with evidence for broad absorption lines (BALs): DMS0100.1-0058 (z=1.73), and DMS2139-0351 (z=1.91). DMS1358-0045 might also be a BALQSO, but a higher S/N spectrum is needed to determine this and to secure its redshift. The number of BALs is roughly consistent with their 10% expected incidence in optically selected samples. The quasar DMS1358-0055 (z=3.38) has a strong associated CIV absorption system, and DMS1714+4951 (z=3.3) may also show such absorption. Slit spectra with higher S/N ratio and broader wavelength coverage are needed for further study of these objects.

The only z>4 quasar discovered in our survey so far is DMS2247-0209. A finding chart is shown in Figure 4, the HYDRA spectrum from the discovery run is shown in Figure 5, and a spectrum extending farther to the red, obtained by J. Ge and J. Bechtold at the MMT on 1993 Oct. 20, is shown in Figure 6. The strong emission line is recognizable as Lyα+NV, with the Lyman-$\alpha$ forest reducing the flux blueward of the line. We measure flux deficit parameters (Schneider, Schmidt, & Gunn 1989) $D_A$=0.76±0.13 and $D_B$=0.86±0.08, a bit high even for z>4 quasars (Kennefick et al. 1995), but these values are very dependent on the highly uncertain placement of the continuum. There is a probable damped-Ly$\alpha$ system along the line of sight at z=4.11±0.01. We detect Ly$\beta$ and possibly CIV 1549 absorption from this system, as well as a Lyman-limit system at z=4.14±0.01 which is likely caused by the red wing of the damped system.

Despite the identification of several emission lines, the exact redshift of DMS2247-0209 is uncertain. The spectrum shows either weak BAL or strong associated-absorption features (or possibly both) at z=4.21 and z=4.25 in CIV 1549, SiIV/OIV] 1400, NV, and Ly$\alpha$. If the strong emission line at 6580Å is predominantly Ly$\alpha$, then z=4.41±0.02 and the red shoulder of the line is due to NV; but in this case the expected Ly$\beta$, SiIV/OIV], and CIV peaks are slightly redward of their apparent peaks. If the line is instead predominantly NV, then z=4.33±0.02 and the expected Ly$\beta$, SiIV/OIV] and CIV peaks align with the observed peaks, but we must then assume substantially more BAL and/or associated absorption of CIV and Ly$\alpha$ relative to NV and SiIV/OIV] in order to explain the observed line ratios. The line has an observed EW$\sim$200Å, which is consistent with either very strong NV or weak or strongly absorbed Ly$\alpha$. We assume z=4.33 since Ly$\alpha$ is likely to suffer the most absorption and is thus the least reliable line for redshift determination, but we cannot rule out z=4.41. An exact redshift determination could probably best be done with IR spectroscopy of the rest-frame optical narrow emission lines.

### 4.2. Compact Narrow Emission-Line Galaxies

A major source of contamination in simple UVX-selection of quasars is low-redshift compact (i.e., unresolved) narrow emission-line galaxies (CNELGs). Table 2 lists positions, redshifts, magnitudes, and notes on the CNELGs discovered in our survey. The redshift distribution of our CNELGs is given in Figure 7; all but two are at z<0.55. We find that CNELGs are usually not as blue in U-B or in B-V as quasars (see Figure 1a), similar to the results of Koo & Kron (1988; their Figure 2). For our survey, a simple requirement of B-V$\leq$0.60 suffices to eliminate $\sim$65% of the UVX-selected CNELGs without eliminating any known UVX quasars. These galaxies are interesting in their own right, however, since they consititute a population of galaxies overlooked in morphologically-selected samples of faint galaxies. The redshift distribution is as expected from such surveys, but the emission-line properties of the two samples may be different (Smetanka 1992).

### 4.3. Stars

The spectroscopically confirmed stars have so far been only roughly classified, but among the unremarkable contaminating normal stars we have tentatively identified several DA white dwarfs, one DB white dwarf, a number of hot metal-weak stars, a few metal rich stars, and a few late M dwarfs with colors and spectra consistent with being M6 or M7 based in the classification scheme of Kirkpatrick et al. (1991).

### 4.4. Comparison with the Literature

The NASA/IPAC Extragalactic Database (NED) was used to search the literature for objects from the literature located within our fields. The one previously known quasar in our fields, 1714+502 (z=1.129), was independently recovered by our survey (DMS1714+5016). In May 1993 a spare fiber was used to obtain a spectrum of IRAS FSC 17140+5011, which was easily identified on the images as a morphologically disturbed galaxy, and turned out to have z=0.232.

## 5. Synthetic Quasar Spectra and Colors

To determine the regions of multicolor space most likely to be populated by quasars, and to help devise selection criteria efficient at selecting quasars, we generated synthetic quasar spectra and computed their colors in our filter set. The simulations cover a wide range of quasar SEDs and accurately account for the effects of intergalactic HI absorption. We believe that they satisfactorily represent the range of colors exhibited by optically-selected quasars. We can thus use them to devise efficient selection criteria and to calculate accurate detection probabilities for quasars of various redshifts and SEDs. Knowledge of these detection probabilities allows us to make accurate incompleteness corrections, which is essential for accurate determination of the luminosity function.

Our simulated quasars are essentially the same as those described in Warren et al. (1991), with the slight modifications described in Warren, Hewett, & Osmer (1994; hereafter referred to as WHO). They consist of nine 'spectral types' defined by three different values of the Ly$\alpha$+NV emission-line strength (42, 84, and 168 Å in the rest frame) and three values of the spectral index $\alpha$ (-0.25, -0.75, and -1.25), which is the slope of the power-law continuum $f_\nu \propto \nu^\alpha$. Note that emission line ratios were fixed relative to the Ly$\alpha$/NV line, and only the strength was allowed to vary. Simulated spectra of each spectral type are generated in redshift bins spaced every 0.1 in redshift from z=0.05 to z=6.45. We include the effects of photometric errors and galactic extinction, but the images of our fields were all taken within at most four days of each other, so we ignore variability.

At z>2.2, the effects of intergalactic HI absorption blueward of Ly$\alpha$ were simulated by creating 100 synthetic transmission functions at each z, following the prescription of WHO and based on the work of Møller & Jakobsen (1990). These transmission functions account for absorption from the first 17 Lyman series transitions as well as continuum absorption below the Lyman limit from three populations of absorbing clouds: Lyman-forest, Lyman-limit, and damped-Ly$\alpha$ systems.

Details of the column density and redshift distribution of the model absorbers can be found in WHO, appendix A. At z>5 the absorber parameters are extrapolations from lower-redshift data. In the range 1.5<z<2.2, intergalactic HI absorption affects only the U band, so we modeled it with a simple depression of the continuum blueward of Lyman-$\alpha$, according to the prescription of Warren et al. (1991). Again, 100 separate spectra were created, each with a different, random value of the flux deficit. At z<2.2, we also extended the simulated quasars of WHO by adding weaker and longer-wavelength emission lines. These simulations include all the lines from Wilkes (1986) plus H$\alpha$. Equivalent widths were taken from Wilkes (1986) except for OVI (Shuder & MacAlpine 1979) and H$\alpha$ (Cristiani & Vio 1990).

Thus for each redshift bin with z>1.5, we have 900 synthetic quasar spectra. At z<1.5, there are only nine per redshift bin, since the Lyman-$\alpha$ forest does not affect observed optical wavelengths at those redshifts. From these spectra we synthesize instrumental magnitudes for each passband by convolving the quasar flux density per unit wavelength $F_\lambda$ with each filter's sensitivity function $S(\lambda)$. $S(\lambda)$ is the fraction of photons of wavelength $\lambda$ that pass through the system, taking into account one reflection from aluminum, the transmission of the doublet corrector, the transmission of the filter, and the quantum efficiency of the CCD. It is normalized such that $\int S(\lambda) d\lambda = 1$.

Galactic extinction was incorporated into the simulated quasar magnitudes by convolving the galactic extinction curve of Howarth (1983) with the normalized filter response functions. These magnitude corrections were then scaled by E(B-V) and applied to the simulated instrumental magnitudes. Since E(B-V) is different in each CCD field (see Table 1 of Paper I), we used the area-weighted average of the individual fields' E(B-V) values. This value of E(B-V)=0.0269 leads to absorption values of 0.129, 0.112, 0.085, 0.069, 0.058, and 0.047 magnitudes, respectively, for the U, B, V, R, I75, and I86 filters.

The simulated magnitudes were shifted to magnitudes spaced every $0.1^m$ in either B, V, or R, depending on the selection criteria being considered, and ranging from the saturation limit to the faintest magnitude bin from which candidates were deemed to be reliably selectable. Photometric errors were then simulated in every filter by choosing randomly from a gaussian distribution scaled to the observed photometric error at each magnitude. This very accurately mimics the uncertainty in observed magnitudes and colors at faint magnitudes in the survey.

The instrumental magnitudes were next treated as observations made at zero airmass and photometrically calibrated using essentially the same parameters used for calibration of our objects. The exact parameters of the calibration were established as follows. We synthesized colors of a variety of main sequence stars taken from the digital library of Silva & Cornell (1992), using as the initial zeropoint for each filter the values listed in Table 4 of Paper 1. To bring the synthetic stellar main sequence into line with the observed one in a variety of color-color diagrams, adjustments of $0.05^m$ and $0.15^m$ were required in the zeropoints of the I75 and I86 filters respectively. Adjustments at this level remain within the range of those made for the CCD fields themselves and thus are not distressing. Finally, just as in the real observations, two simulated measurements were made for each object and combined in exactly the same manner as the real data. This was done using the average magnitude limits for the entire survey.

The final synthesized quasar colors match the observed quasar colors quite well. Figure 8 is a U-B/B-V color-color diagram where simulated quasars at z<3 and B<21 are plotted as points and observed z<3 quasars as squares. The observed quasar colors fall within the areas covered by the simulations, with a few minor exceptions. One quasar has a much bluer B-V color than expected (DMS1715+4952, z=1.85, U-B=-0.778, B-V=-0.425). This object's spectrum shows a rather blue continuum and an abnormally high CIV/CIII] ratio, which is probably sufficient to explain its colors since CIV 1549 is in B and CIII] 1909 in V at this object's redshift. (Recall that our synthetic spectra all have identical line ratios.) Several quasars at z∼1 are clumped at the red end of the simulated colors for objects at that redshift, at U-B∼-0.8, B-V∼0.5. (This effect can also be seen in U-V/V-R and B-V/V-R diagrams; see Figures 11a and 11b.) This might again be an artifact of our assumption of fixed line ratios, since MgII 2798 is in V at z∼1. Also, we have not included Fe II or Balmer continuum emission in our synthetic spectra, which can contribute considerable flux at rest wavelengths >2200Å (Francis et al. 1991), which would also redden the observed B-V and V-R colors. Many of these objects' spectra do indeed show strong MgII and/or FeII emission; the rest have spectra of S/N too low to say for sure. Thus our synthetic quasars adequately mimic the range of colors observed in real quasars, with the only minor exceptions arising from effects knowingly neglected in the models; namely, line ratio variations and FeII emission. Finally, we note that while the range of allowed quasar colors is important, the distribution of quasars within that range is also important. We have not determined this distribution, since it is difficult to estimate for the simulated colors at z<1.6 where we have only 9 simulated colors at each redshift (excluding photometric errors). In addition, any discrepancy we might find between observed and simulated colors would be of questionable significance due to the small number of quasars in our sample.

## 6. Refinement of Selection Criteria

We are now able to refine our selection methods and determine their efficiencies. The spectroscopically identified objects can be used to improve any selection method by studying its *observational efficiency*: the fraction of quasars found among the spectroscopically identified objects selected as candidates by the method. For example, we mentioned in §4.2 that an additional criterion of B-V<0.6 greatly lessens the contamination from galaxies in a UVX-selected sample. Also, by finding the percentage of simulated quasars in a given redshift and apparent magnitude bin selected by each method, we can estimate the *detection probability* for such quasars. In the absence of other limiting factors, and assuming that our simulations adequately describe the true color distribution of quasars, the detection probability is our completeness.

We consider here three selection methods: a modified UVX selection, U-V/V-R + B-V/V-R diagram selection (which for simplicity we refer to as VRX selection even though it is not based on a V-R excess), and BRX selection. These methods are aimed roughly at z<3, 3<z<4, and z>4, respectively. We consider candidate selection down to fainter magnitudes than we have currently reached spectroscopically, but even at such magnitudes the completeness of the catalog is no less than 93%. The total fraction of misclassified extended objects in the catalog to B=23.1, V=22.0, and R=22.0 is estimated to be 7.5%, 4.5%, and 10% respectively. See Paper I, §4.3 for discussion of how these estimates were made.

In all the color-color diagrams we consider, we drew up simple selection criteria by following a few guiding principles:

- maximize the fraction of simulated quasar colors selected by the criteria

- include as many spectroscopically confirmed quasars as feasible and exclude as many spectroscopically confirmed stars and CNELGs as possible within the selection region

- probe as close to the stellar locus as possible in each magnitude bin, but not so close that a large number of objects very close to the locus are included by the criteria

This last guideline is an important one, but has only been followed subjectively; how close to place the selection criteria to the stellar locus was decided purely by eye. Ideally one would like to simulate the broadening of the stellar locus with magnitude and develop criteria that exclude objects close enough to the locus that they could be stars at, for example, the $4\sigma$ level at each magnitude. This could be done by considering the locus as a set of points in color-color space, convolved with a gaussian error distribution as a function of magnitude and color.

Figures 9-15 present color-color diagrams for the various selection methods. See §2.1 and the caption to Figure 9 for a thorough guide to these diagrams. Not all spurious outliers have been removed from the catalog used to make these plots, so some of the 'obvious' candidates without spectroscopic IDs are in fact spurious. Also, the spectroscopically confirmed quasars and CNELGs are plotted in all diagrams regardless of their magnitudes. Finally, the thick lines denote regions of color-color space from which we select quasar candidates for spectroscopic followup.

### 6.1. UVX selection

In Figure 9, we plot the U-B/B-V diagram for catalog objects in the magnitude bins 16.5<B<21 and 21<B<22. In Figure 10, we plot the UVX portion of the U-B/B-V diagram for catalog objects in the magnitude bins 22<B<22.6 and 22.6<B<23.1. It is easy to see from Figure 9a that the standard UVX criterion of U-B<-0.3 selects large numbers of both confirmed and simulated quasars, but that there is a population of quasars (typically at z∼2.5) with U-B>-0.3 and B-V<0.3 that inhabit a region also occupied by A stars. Since A stars are rare at high galactic latitude, we can include such objects without adding greatly to the contamination. Because we do so, our selection of objects from the U-B/B-V diagram is not strictly a UVX criterion; however, we retain that nomenclature for convenience.

We adopt the following selection criteria for the U-B/B-V diagrams.

- At magnitudes 16.5<B<21, we require B-V<0.35 for U-B>-0.3 and B-V<0.6 for U-B<-0.3. 22 of the 38 spectroscopic IDs among these candidates are found to be quasars, for an observational efficiency of 58%.

- At magnitudes 21<B<22, we require B-V<0.30 for U-B<-0.4 and B-V<0.6 for U-B<-0.4. 15 of the 26 spectroscopic IDs among these candidates are found to be quasars, for an observational efficiency of 57%.

- At magnitudes 22<B<22.6, we require B-V<0.25 for U-B>-0.5 and B-V<0.6 for U-B<-0.5. 2 of the 10 spectroscopic IDs among these candidates are found to be quasars, for an observational efficiency of 20%.

- At magnitudes 22.6<B<23.1, we require B-V<0.4 and U-B<-0.5. At these faint magnitudes there is considerable broadening of the stellar locus and contamination by misclassified galaxies. However, it is still possible to construct selection criteria reasonably efficient at selecting a large percentage of simulated low-redshift quasars, even if the observational efficiency is almost certainly lowered. Since we have no spectroscopic IDs of candidates this faint, the observational efficiency cannot be estimated for this magnitude bin.

The total numbers of candidates selected from the catalog by these criteria are 79, 68, 99, and 145 respectively.

### 6.2. U-V/V-R + B-V/V-R selection

Figure 11a-b shows the U-V/V-R and B-V/V-R diagrams for the magnitude bin 17<V<20.5. Figure 12a-b shows the U-V/V-R and B-V/V-R diagrams for the magnitude bin 20.5<V<22. The simulated quasar colors show that at redshifts z>3 many quasars are red enough in U-V to distance themselves from the stellar locus in the U-V/V-R diagram, while in the B-V/V-R diagram the colors swing back toward the locus at z~4 after initially moving away from it at z~3. In both these diagrams, in the fainter magnitude bins, the wider stellar locus encroaches considerably on the outlier regions where quasars can be found, so we opted to combine selection criteria from the two diagrams and require that objects be located in outlying areas of both diagrams. This helps reduce accidental stellar contamination of the outlying areas and improves the selection efficiency noticably.

We adopt the following selection criteria for the U-V/V-R + B-V/V-R selection, requiring objects to be in certain regions of each diagram:

- At magnitudes 17<V<20.5, we require in the U-V/V-R diagram U-V>0.4 for V-R<0.26, U-V>5.21×[(V-R)]-0.95 for 0.26<V-R<0.84, and U-V>3.40 for V-R>0.84. We also require, in the B-V/V-R diagram: B-V>0.5 for V-R<0.25, B-V>1.8×[(V-R)]+0.05 for 0.25<V-R<0.97, and B-V>1.8 for V-R>0.97. The only spectroscopic ID made so far among the 3 candidates selected this way was found to be a quasar, for an observational efficiency of 100%.

• At magnitudes 20.5<V<22, we require in the U-V/V-R diagram U-V>1.0 for V-R<0.23, U-V>3.08×[(V-R)]+0.30 for 0.23<V-R<0.86, and U-V>2.95 for V-R>0.86. We also require, in the B-V/V-R diagram: B-V>1 for V-R<0.35, B-V>1.85×[(V-R)]+0.35 for 0.35<V-R<0.89 B-V>2 for V-R>0.89. The only two spectroscopic IDs made so far among the 3 candidates selected this way were both found to be quasars, for an observational efficiency of 100%.

### 6.3. BRX selection

Figure 13a-b shows the B-R/R-I86 diagram for two bright magnitude bins, with limiting magnitudes of R=20 and R=21 respectively. Figure 14a-b shows the B-R/R-I86 and B-R/I75-I86 diagrams for the magnitude bin 21<R<21.5. Figure 15a-b shows the B-R/R-I86 and B-R/I75-I86 diagrams for the magnitude bin 21.5<R<22.

Quasars with $z \gtrsim 3$ are easily seen to be redder in B-R and bluer in R-I86 or I75-I86 than most stars. We adopt the following criteria for selection:

• At magnitudes 17.5<R<20, we require B-R>0.5 for R-I86<-0.32, B-R>2.65×(R-I86)+1.35 for -0.32<R-I86<0.77, and B-R>3.4 for R-I86>0.77. 1 of the 4 spectroscopic IDs among these candidates was found to be a quasar. However, the redshift of this object (DMS0100.3-0058) is z=1.17, which is not predicted to be selected by this method. DMS0100.3-0058 is one of the z~1 quasars suspected of having strong FeII emission based on its position in the U-B/B-V diagram (see §5). At that redshift, substantial FeII emission appears in the R band, giving the objects red B-R and blue R-I86 colors (in this case, 0.81 and -0.23 respectively). These colors allow DMS0100.3-0058 to be selected since it is a bright object; at fainter magnitudes it would not have been selected. The observational efficiency for high-redshift quasars is thus 0%.

• At magnitudes 20<R<21, we require B-R>0.5 for R-I86<-0.36, B-R>2.75×(R-I86)+1.5 for -0.36<R-I86<0.8, and B-R>3.7 for R-I86>0.8. 3 of the 6 spectroscopic IDs among these candidates were found to be quasars, for an observational efficiency of 50%.

The total numbers of candidates selected from the catalog by these criteria are 7 and 16, respectively.

At R>21, because of the broadening of the stellar locus and the presence of large numbers of faint red stars for which only limits on the B-R color can be set, many regions of interest on the color-color diagrams are not accessible. Still, some regions are accessible, and by requiring that candidates meet color criteria in both the B-R/R-I86 and B-R/I75-I86 diagrams we can help reduce accidental stellar contamination. Thus we require objects to be found in certain regions of each diagram:

• At magnitudes 21<R<21.5, we require in the B-R/R-I75 diagram B-R>1.35 for R-I86<-0.125 and B-R>3.21×(R-I86)+1.75 for R-I86>-0.125. We also require in the B-R/I75-I86 diagram: B-R>1.35 for I75-I86<-0.09 and B-R>4.4×(I75-I86)+1.75 for I75-I86>-0.09.

• At magnitudes 21.5<R<22, we require in the B-R/R-I75 diagram B-R>1.8 for R-I86<-0.1 and B-R>2.6×(R-I86)+2.05 for R-I86>-0.1. We also require in the B-R/I75-I86 diagram: B-R>1.8 for I75-I86<-0.25 and B-R>2.58×(I75-I86)+2.45 for I75-I86>-0.25.

The total numbers of candidates selected from the catalog by these criteria are 2 and 3, respectively.

No estimate of the observational efficiency can be made for these R>21 magnitude bins because no observed candidate from these bins has yet been spectroscopically identified. We assume an efficiency of 30% based on the ∼30% total efficiency achieved at brighter magnitudes with the B-R/R-I86 selection.

## 7. Detection Probabilities and Incompleteness Corrections

The simulated quasar colors were subjected to the same selection criteria as the real objects, in order to calculate the detection probabilities as a function of quasar spectral type and redshift. We calculated the detection probabilities in redshift bins spaced every 0.1 in z from z=0.05 to 6.45 and magnitude bins spaced every $0.1^m$ in either B, V, or R. The detection probability for each spectral type was calculated separately, since the spectral types are not equally common. The total detection probability for a given selection criterion in each redshift bin is the detection probability for each spectral type times the fraction of quasars having that spectral type. The relative numbers of each spectral type were taken from Table 7 of WHO. (Strictly speaking, those numbers are for a survey limited on continuum magnitude; however, the relative numbers in a survey like ours, limited on apparent magnitude, were calculated and found to be not significantly different.)

Lastly, the detection probabilities were corrected for incompleteness. In Paper I it was shown that the catalog is 100% complete down to B=21.5, V=21, and R=20 and then declines linearly to 90% completeness at the $5\sigma$ limiting magnitudes of B=23.8, V=23.5, and R=23 (Table 3 of Paper I). The exact completeness factors were calculated for each magnitude bin and multiplied into the detection probability for that bin. This is a small effect; even at the faintest magnitudes where candidates were selected, the catalog is never less than 93% complete.

These detection probabilities are plotted in Figures 16a, 16b, and 16c for the UVX, U-V/V-R + B-V/V-R, and BRX selection criteria, respectively. Contour levels are 0, 20, 40, 60, 80, and 95%. Our three criteria complement each other rather well, with the redshift of lowest detection probability being z∼3. The steplike structure seen toward faint magnitudes is due to the abrupt change in selection criteria at specific magnitudes, rather than gradually. The criteria at faint magnitudes are less efficient because they cover a smaller area in the color-color diagrams to avoid the spreading stellar locus. However, within a given 'step', the efficiency at a given redshift actually increases slightly with magnitude because the increasing errors scatter more objects into the selection areas. The competing effect not represented on these diagrams is that stellar contamination will also increase toward fainter magnitudes within a given 'step'. The abrupt dropoffs seen in the VRX contour plot at V∼20, z>4 and in the BRX contour plot at R∼20.6, z>5 are due to the U-V and B-R color limits for these faint objects becoming bluer than the selection criteria. For example, at R>20.6, a z>5 quasar will have extremely red B-R and R-I86 because Ly$\alpha$ has moved through the R bandpass. But the average $3\sigma$ limiting B magnitude is 24.3, so the object will have B-R<3.7, which is just bluer than the BRX selection criteria. In reality, because the limiting magnitudes vary from field to field, these abrupt dropoffs will be smoothed out by perhaps ±0.1 magnitude. The slight dip in the UVX contour plot at B∼22, z∼0.7 is due to the well-known reddening caused by MgII 2798 contributing flux in the B band at that redshift while CIII] 1909 contributes little in the U band.

## 8. Comparison of Candidate Counts to Quasar Surface Density Predictions

Spectroscopic followup of our quasar candidates is not yet complete, so a full derivation of the luminosity function would be premature. However, the data in hand allow us to compute predicted quasar counts using both observational and theoretical approaches, and to compare the two predictions. The observational approach predicts quasar counts by multiplying the number of candidate objects we have selected using a given method by the *observational efficiency* for that method (the number of confirmed quasars divided by the number of spectroscopic IDs). If the extrapolation of the observational efficiency to fainter magnitudes is correct, and if the tendency to observe the best candidates first is accounted for, this method should yield an accurate empirical estimate of quasar counts. The theoretical approach predicts quasar counts by integrating the quasar luminosity function (QLF) determined from previous work to give a prediction of quasar surface densities. Multiplying this number by the detection probability for a given selection method gives the number of these quasars that should be detected in our survey. If these detection probabilities have been determined accurately, this method gives an independent, theoretical prediction of quasar counts based on earlier work on the QLF. Comparison of the counts predicted by these two methods will give a preliminary check on how the results of this survey compare to previous work.

For the 'theoretical' prediction of quasar number counts at z<2.3, we use the results of Koo & Kron (1988; hereafter KK88); specifically, their fit to their observed number counts rather than the full luminosity function. The results of KK88 are in good agreement with other faint UVX quasar surveys such as Boyle, Shanks, & Peterson (1988); Boyle, Jones, & Shanks (1991); and Zitelli et al. (1992). At z>3 we use the results of WHO (Warren, Hewett, & Osmer 1994) primarily for convenience, as it was a red-selected sample. The more recent results from the Palomar Transit

Grism Survey of Schmidt, Schneider, & Gunn (1995) and the Digital Palomar Sky Survey of Kennefick, Djorgovski, & de Carvalho (1995) are in good agreement with the results of WHO. The best-fit QLF model of WHO is the double power-law pure luminosity evolution (PLE) exponential model of Boyle, Shanks, & Peterson (1988), modified such that only the bright power law evolves. This model parameterizes the luminosity function $\Phi$ by a function of the form

$$\Phi(M, z)\, dM\, dV = \frac{\Phi^*}{[10^{0.4(M-M(z))(\alpha+1)} + 10^{0.4(M-M^*)(\beta+1)}]}\, dM\, dV \qquad (3)$$

where

$$M(z) = M^* - 1.086\, k_L \tau \qquad (4)$$

and the best-fit values for the parameters (for $q_o=0.5$) are $\alpha=-5.05$, $\beta=-2.06$, $M^*=-13.21$, $k_L=10.13$, and $\Phi^*=0.102$ Mpc$^{-3}$ mag$^{-1}$. WHO parameterize the observed decline in space density at $z>3.3$ with a pure density evolution (PDE) model wherein the QLF maintains the same shape at $z>3.3$ as at $z=3.3$, but whose amplitude is lower by the factor $\exp[(3.3-z)k_D]$, with $k_D=3.33\pm1$ for $q_o=0.5$. The WHO QLF is used to find the surface density in a given apparent magnitude and redshift range via a double integral. First, the luminosity function is integrated over the absolute magnitude limits corresponding to the apparent magnitude limits at each redshift to obtain the total observed space density at that redshift. Then this space density is integrated over the volume intercepted by the survey area in the given redshift range, yielding a surface density.

Using the data from KK88 and WHO, we calculate the expected number of quasars in the 0.83□° area of our survey in redshift and apparent magnitude bins of width $\Delta z=\Delta m=0.1$. Multiplying these numbers by the detection probability for a given selection method in each bin, we obtain the theoretically predicted number of quasars among the candidates selected by that method. Summing these numbers over the appropriate redshift and magnitude ranges, we can compare them to the observationally predicted number of quasars, which are given by the observed number of candidates multiplied by the observational efficiencies given for each selection method in §6.

First we consider the UVX-selected candidates. For each magnitude bin, Table 3 lists the number of good candidates, the number of spectroscopic IDs, the observed numbers of $z<2.3$ and $z>2.3$ quasars, and the total observational efficiency, including $\pm 1\sigma$ confidence ranges. The observational efficiency is used to estimate the total number of $z<2.3$ and $z>2.3$ quasars among the candidates in each magnitude bin. In the last two columns we give the number of $z<2.3$ and $z>2.3$ quasars we expect to find among our candidates. These numbers are generated from the number-magnitude relation of KK88, scaled to the area of our survey, and convolved with the UVX detection probabilities given in §6.1.

We first note that the number of candidates is safely larger than the predicted number of quasars in each magnitude bin. At B<21, we have spectroscopically identified less than half of our candidates and yet we have already identified almost as many quasars as KK88 predict for the entire magnitude bin. Between B=21 and B=22, our estimate of the number of quasars among our candidates based on spectroscopic IDs to date is exactly in line with the predictions of KK88. Between B=22 and B=22.6, we have only 10 spectroscopic IDs and only 2 quasars so far out of 99 candidates. At B>22.6, we have no spectroscopic information to date.

Thus our observations match KK88 well at intermediate magnitudes and are entirely consistent at faint magnitudes, but show an excess at bright (B<21) magnitudes and low (z<2.3) redshifts. Is there any reason to expect our remaining B<21 candidates contain a smaller percentage of quasars than the identified candidates? The remaining candidates are not strongly clustered in particular fields or near the upper or lower magnitude limits of the magnitude bin. However, as can be seen in Figure 9a, a larger fraction of them occupy the region U-B>-0.3 and B-V<0.35, where we expect greater contamination from A stars. Calculating the observational efficiency for this region separately from the rest of the UVX selection region, we estimate a total of 17 z<2.3 quasars remaining among our unidentified bright UVX candidates, instead of the 22.5 estimated using only one observational efficiency for the entire UVX region. Thus, while this refinement predicts ∼36 instead of ∼41 z<2.3 quasars among our B<21 UVX candidates, that is still larger than the 22 predicted by KK88. Quantitatively, the detection of 36 quasars in a sample in which we expect 22 would be a $3\sigma$ excess, using the Poisson $3\sigma$ limits for small numbers of events tabulated in Gehrels (1986). Alternatively, if the selection techniques used in KK88 are ∼30% incomplete as suggested by Majewski et al. (1993), then the corrected KK88 counts to B=21 are in excellent agreement with our estimated counts to B=21. However, this correction to the KK88 counts would also make our counts at B>21 30% incomplete, which we feel is unlikely. More spectroscopic IDs are needed to resolve this issue.

In Table 4 we consider the VRX and BRX selection methods for various magnitude ranges. For each magnitude range, we give the number of candidates selected according to the criteria of §6.2 or §6.3, the number of objects spectroscopically identified, the number of spectroscopically confirmed quasars with z>3, and the observational efficiency (the ratio of the previous two entries). The next column is the estimated number of z>3 quasars among the candidates, which is just the number of candidates times the observational efficiency. The final column shows the predicted number of candidates according to the QLF model of WHO. Note that the observed and predicted quasars detectable via each method are not necessarily independent objects, as some quasars are detectable via both the VRX and BRX methods.

Considering first the VRX method, we see that we have spectroscopically identified 3 of this method's 6 candidates and that 3 of our 4 z>3 quasars are among the candidates. The WHO QLF model predicts 5.8 quasars detectable by this method with 17.5<V<22, so our results are consistent with expectations.

Considering now the BRX method, we adopt an observational efficiency of 30% at all magnitudes, based on the identification of 3 z>3 quasars among our 10 spectroscopic IDs of candidates selected via this method. The WHO QLF model predicts about two z>3 quasars should be detectable by us to R=22.0. However, we have observed only 36% of the BRX candidates to date and have already identified 3 z>3 quasars among them. Is there any reason to expect our remaining BRX candidates contain a smaller percentage of quasars than the identified candidates? The remaining candidates are not strongly clustered in particular fields or near the upper or lower magnitude limits of the magnitude bins, nor do they occupy different regions of color-color space than the spectroscopically identified BRX candidates. Contamination from the stellar locus could increase at fainter magnitudes, but the fact that we have only five candidates at R>21 argues that our selection criteria are quite restrictive. Contamination from misclassified galaxies does increase at fainter magnitudes, but since they comprise only 10% of the catalog entries by R=22 only ~3 of our BRX candidates are likely to be misclassified galaxies.

Thus there is no compelling reason to believe our remaining BRX candidates harbor a smaller percentage of quasars than the identified candidates. There is only a 5.2% chance that the three z>3 quasars we have found so far are the only ones among the remaining 25 good BRX candidates, based on the probability of those specific three objects being among a random sample of 10 objects chosen from the total sample of 25. Our best estimate is that a total of about eight quasars with z>3 are in our BRX-selected sample. The confirmed detection of eight quasars in a sample where two are expected would be just over a $3\sigma$ excess (Gehrels 1986), and our highest priority for follow-up spectroscopy is to observe the remaining BRX candidates.

## 9. Summary

We have presented initial results from a multicolor quasar survey. We have devised various methods of multicolor selection of quasar candidates from a catalog of 21,375 stellar objects. From spectroscopic observations we have identified more than 40 quasars, including one at z=4.33. We have also identified 39 compact narrow emission-line galaxies. We have determined the efficiency of the selection methods via extensive and detailed simulations of quasar colors. We find that quasars can be selected to z=3 from various regions of the U-B/B-V diagram, but the efficiency drops rapidly at z>2.6. At z>3 the U-V/V-R + B-V/V-R and B-R/R-I86 + B-R/I75-I86 diagrams can be used, leaving only a narrow range 2.7<z<3.2 in which quasars have stellar colors and multicolor selection is inefficient. We have compared the numbers of our candidates and spectroscopic IDs to the number of quasars we expect to detect in our survey area based on the results of earlier work in the field. The agreement of our observations with these expectations is good in most cases. However, we do estimate that our survey contains more quasars with B<21 and z<2.3 than expected from the results of Koo & Kron (1988) and more z>3 quasars than expected from the results of Warren, Hewett, & Osmer (1994), both at the $3\sigma$ level. More spectroscopic work is needed to confirm the reality of these estimates, and indeed spectroscopy of our remaining faint quasar candidates is continuing. In Paper III (Osmer et al. 1996) we will present final spectroscopic results and a full derivation of the luminosity function.

We wish to thank many people for their assistance and help during the course of this project, including Frank Valdes for essential help with the HYDRA reductions, Taft Armandroff, Sam

Barden, and the rest of the HYDRA team for building and maintaining the instrument and for helping us to use it efficiently, the KPNO mountain staff, particularly the telescope operators, for their assistance in observing, and last but definitely not least the KPNO TAC for their allocation of time for this project. PBH acknowledges support from an NSF Graduate Fellowship and a University of Arizona Graduate College Fellowship. This research has made use of the NASA/IPAC Extragalactic Database (NED) which is operated by the Jet Propulsion Laboratory, California Institute of Technology, under contract with the National Aeronautics and Space Administration.

Table 1. Quasars: Identifications

| No. | I.D. | R.A.(1950) | Dec.(1950) | B mag | z | Notes |
|---|---|---|---|---|---|---|
| 1 | DMS0059−0051 | 00:59:04.11 | −00:51:30.13 | 20.7 | 0.84? | 1 line, MgII? |
| 2 | DMS0059−0056 | 00:59:09.40 | −00:56:08.91 | 21.4 | 2.63 | |
| 3 | DMS0059−0103 | 00:59:12.19 | −01:03:09.86 | 21.3 | 0.71? | 1 line, MgII? |
| 4 | DMS0059−0059 | 00:59:37.85 | −00:59:50.00 | 20.5 | 1.92 | |
| 5 | DMS0100.1−0058 | 01:00:07.54 | −00:58:14.62 | 19.4 | 1.73 | BAL |
| 6 | DMS0100.3−0058 | 01:00:18.76 | −00:58:09.87 | 20.8 | 1.17 | |
| 7 | DMS0100−0057 | 01:00:38.12 | −00:57:43.73 | 20.6 | 0.77? | 1 line, MgII? |
| 8 | DMS1358−0039 | 13:58:09.82 | −00:39:41.26 | 21.5 | 1.34? | CIII]?, MgII? |
| 9 | DMS1358−0047 | 13:58:17.46 | −00:47:43.01 | 20.8 | 2.04 | |
| 10 | DMS1358−0054 | 13:58:27.83 | −00:54:16.19 | 20.4 | 2.81 | |
| 11 | DMS1358−0036 | 13:58:29.13 | −00:36:02.66 | 18.8 | 0.9? | 1 line, MgII? |
| 12 | DMS1358−0100 | 13:58:36.64 | −01:00:51.78 | 20.9 | ? | UVX, broad lines |
| 13 | DMS1358−0031 | 13:58:40.16 | −00:31:09.44 | 19.1 | 2.55 | |
| 14 | DMS1358−0055 | 13:58:45.78 | −00:55:29.66 | 21.4 | 3.38 | Assoc. absorp. |
| 15 | DMS1358−0050 | 13:58:47.32 | −00:50:23.37 | 21.0 | 1.34 | |
| 16 | DMS1358−0045 | 13:58:55.56 | −00:45:22.88 | 21.0 | ? | BAL? |
| 17 | DMS1359−0030 | 13:59:01.13 | −00:30:50.82 | 21.5 | 0.86 | |
| 18 | DMS1359−0056 | 13:59:05.87 | −00:56:51.58 | 21.5 | 1.85 | Weak lines |
| 19 | DMS1359−0036 | 13:59:11.02 | −00:36:45.15 | 21.3 | 1.14? | 1 line, MgII? |
| 20 | DMS1714+5019 | 17:14:05.37 | +50:19:14.53 | 20.9 | 1.52 | Weak lines |
| 21 | DMS1714+5022 | 17:14:08.58 | +50:22:51.88 | 21.8 | 1.8? | Weak lines |
| 22 | DMS1714+5016 | 17:14:08.72 | +50:16:28.93 | 19.8 | 1.12 | |
| 23 | DMS1714+5014 | 17:14:11.80 | +50:14:56.40 | 22.1 | 1.19 | |
| 24 | DMS1714+4951 | 17:14:13.46 | +49:51:56.09 | 22.7 | 3.3 | Assoc. absorp.? |
| 25 | DMS1714+5002 | 17:14:27.02 | +50:02:51.49 | 20.0 | 1.94 | |
| 26 | DMS1714+5020 | 17:14:49.36 | +50:20:53.19 | 19.8 | 0.68 | |
| 27 | DMS1714+5003 | 17:14:50.27 | +50:03:37.95 | 21.3 | 2.8 | |
| 28 | DMS1714+4959 | 17:14:50.86 | +49:59:17.29 | 21.8 | 1.55? | 1 line, Mg II? |
| 29 | DMS1714+5000 | 17:14:55.53 | +50:00:01.54 | 20.7 | 1.95 | |
| 30 | DMS1715+5015 | 17:15:00.78 | +50:15:10.72 | 20.6 | 1.55 | |
| 31 | DMS1715+5014 | 17:15:11.52 | +50:14:59.56 | 21.6 | 1.62 | |
| 32 | DMS1715+5013 | 17:15:16.94 | +50:13:04.63 | 21.5 | 1.19 | |
| 33 | DMS1715+4952 | 17:15:17.20 | +49:52:00.75 | 21.0 | 1.85 | |
| 34 | DMS1715+5024 | 17:15:27.99 | +50:24:02.03 | 21.9 | 1.68? | Weak lines |
| 35 | DMS2138−0352 | 21:38:38.95 | −03:52:07.81 | 21.3 | 1.23 | |
| 36 | DMS2139−0356 | 21:39:00.51 | −03:56:34.39 | 18.8 | 2.36 | |
| 37 | DMS2139.0−0405 | 21:39:02.76 | −04:05:24.08 | 22.0 | 3.32 | |
| 38 | DMS2139.1−0405 | 21:39:11.88 | −04:05:47.32 | 21.5 | 1.60 | |
| 39 | DMS2139−0351 | 21:39:25.00 | −03:51:35.12 | 21.5 | 1.91 | BAL |
| 40 | DMS2140−0401 | 21:40:21.82 | −04:01:39.40 | 21.5 | 2.05 | |
| 41 | DMS2245−0208 | 22:45:42.35 | −02:08:44.33 | 22.2 | 0.86? | 1 line, MgII? |
| 42 | DMS2245−0205 | 22:45:58.52 | −02:05:55.12 | 20.7 | 2.15 | |
| 43 | DMS2246−0215 | 22:46:17.50 | −02:15:46.42 | 19.8 | 1.38 | |
| 44 | DMS2246−0205 | 22:46:41.24 | −02:05:56.57 | 20.2 | 1.92 | |
| 45 | DMS2247−0201 | 22:47:04.93 | −02:01:45.46 | 21.2 | 1.03 | |
| 46 | DMS2247−0209 | 22:47:17.62 | −02:09:26.00 | 24.0 | 4.3 | |

– 23 –Table 2. Compact Narrow Emission-Line Galaxies: Identifications

| No. | I.D. | R.A.(1950) | Dec.(1950) | B mag | z | Notes |
|---|---|---|---|---|---|---|
| 1 | N0059−0059 | 00:59:08.23 | −00:59:50.30 | 22.0 | 0.257 | Weak lines |
| 2 | N0059−0054 | 00:59:10.10 | −00:54:31.54 | 20.6 | 0.223 | Strong lines |
| 3 | N0059−0055 | 00:59:46.79 | −00:55:55.47 | 21.8 | 0.43 | Weak lines |
| 4 | N0100−0059 | 01:00:04.66 | −00:59:52.32 | 21.8 | 0.191 | Weak lines |
| 5 | N0100−0058 | 01:00:08.07 | −00:58:36.56 | 21.8 | 0.174 | Weak lines |
| 6 | N0100−0049 | 01:00:39.94 | −00:49:33.15 | 22.0 | 0.095 | Weak lines |
| 7 | N1358−0035 | 13:58:14.11 | −00:35:33.44 | 21.8 | 0.17? | 1 line, uncertain |
| 8 | N1358−0055 | 13:58:20.30 | −00:55:03.48 | 22.5 | 0.35 | Weak lines |
| 9 | N1358−0041 | 13:58:24.14 | −00:45:47.95 | 22.0 | 0.17 | Weak lines |
| 10 | N1358−0057 | 13:58:24.88 | −00:57:22.37 | 22.5 | 0.54? | 1 line, uncertain |
| 11 | N1358−0102 | 13:58:55.73 | −01:02:06.51 | 23.0 | 0.34? | Uncertain |
| 12 | N1713+5011 | 17:13:54.46 | +50:11:23.00 | 22.0 | 0.11 | Strong lines |
| 13 | N1714+5006 | 17:14:24.85 | +50:06:50.92 | 22.4 | 0.31 | Weak lines |
| 14 | N1714+5018 | 17:14:26.41 | +50:18:05.02 | 22.7 | 0.3 | Weak lines |
| 15 | N1714+4958 | 17:14:27.12 | +49:58:49.57 | 22.4 | 0.229 | Weak lines |
| 16 | N1714.4+5003 | 17:14:27.59 | +50:03:59.94 | 21.3 | 0.29 | Weak lines |
| 17 | N1714.5+5003 | 17:14:31.76 | +50:03:52.34 | 22.2 | 0.812 | Weak lines |
| 18 | N1714+5007 | 17:14:44.87 | +50:07:03.54 | 21.1 | 0.388 | Weak lines |
| 19 | N1714+5009 | 17:14:54.24 | +50:09:49.23 | 22.5 | 0.40 | Weak lines |
| 20 | N1715+5014 | 17:15:00.22 | +50:14:28.48 | 22.3 | 0.44 | Weak lines |
| 21 | N1715.0+5000 | 17:15:01.76 | +50:00:11.48 | 20.8 | 0.12 | Strong lines |
| 22 | N1715+5011 | 17:15:07.48 | +50:11:56.62 | 21.3 | 0.398 | Weak lines |
| 23 | N1715.4+5000 | 17:15:24.30 | +50:00:02.91 | 22.4 | 0.32 | Weak lines |
| 24 | N1715+4957 | 17:15:28.77 | +49:57:29.91 | 21.8 | 0.18 | Weak lines |
| 25 | N2138−0354 | 21:38:49.01 | −03:54:15.07 | 21.8 | 0.300 | Weak lines |
| 26 | N2138−0401 | 21:38:54.09 | −04:01:55.23 | 22.2 | 0.283 | Weak lines |
| 27 | N2139−0353 | 21:39:04.84 | −03:53:21.59 | 22.2 | 0.225 | Weak lines |
| 28 | N2139−0401 | 21:39:13.86 | −04:01:56.49 | 21.4 | 0.343 | Weak lines |
| 29 | N2139−0400 | 21:39:26.36 | −04:00:29.76 | 21.9 | 0.269 | Weak lines |
| 30 | N2139−0358 | 21:39:32.84 | −03:58:51.17 | 21.0 | 0.518 | Weak lines |
| 31 | N2139−0356 | 21:39:32.95 | −03:56:39.86 | 22.1 | 0.454 | Weak lines |
| 32 | N2139−0403 | 21:39:35.77 | −04:03:25.62 | 22.3 | 0.378 | Weak lines |
| 33 | N2245−0213 | 22:45:56.13 | −02:13:01.31 | 21.7 | 0.377 | Weak lines |
| 34 | N2246−0204 | 22:46:38.43 | −02:04:31.62 | 22.1 | 0.79 | 1 line |
| 35 | N2246−0202 | 22:46:43.12 | −02:02:11.44 | 21.1 | 0.294 | Weak lines |
| 36 | N2246−0207 | 22:46:49.20 | −02:07:50.72 | 21.7 | 0.373 | Weak lines |
| 37 | N2246−0209 | 22:46:49.29 | −02:09:57.82 | 21.6 | 0.48 | Weak lines |
| 38 | N2247−0213 | 22:47:10.34 | −02:13:03.59 | 22.4 | 0.357 | Weak lines |
| 39 | N2247−0210 | 22:47:29.23 | −02:10:09.81 | 21.6 | 0.394 | Weak lines |

Table 3. Observed vs. Predicted Numbers of UVX Quasars

| Magnitude Bin Limits | Number of Candidates | Number of Spectroscopic Identifications | Confirmed Quasars (z<2.3) | Confirmed Quasars (z>2.3) | Observational Efficiency (Percent) (z<2.3) | Observational Efficiency (Percent) (z>2.3) | Estimated # of Quasars (z<2.3) | Estimated # of Quasars (z>2.3) | Expected # of Quasars[a] (z<2.3) | Expected # of Quasars[a] (z>2.3) |
|---|---|---|---|---|---|---|---|---|---|---|
| 16.5<B<21.0 | 83 | 38 | 19 | 3 | $50.0\pm8.0$ | $7.9^{+6.7}_{-2.8}$ | $41.5\pm4.9$ | $6.6^{+3.5}_{-2.2}$ | 22.0 | 2.5 |
| 21.0<B<22.0 | 71 | 26 | 14 | 1 | $53.8^{+9.3}_{-9.9}$ | $3.8^{+7.5}_{-0.7}$ | $38.2^{+5.4}_{-5.6}$ | $2.7^{+3.6}_{-1.3}$ | 41.5 | 3.6 |
| 22.0<B<22.6 | 99 | 10 | 2 | 0 | $20^{+17.2}_{-7.8}$ | $0.0^{+16.8}_{-0.0}$ | $19.8^{+15.8}_{-7.9}$ | $0.0^{+15.0}_{-0.0}$ | 46.5 | 3.1 |
| 22.6<B<23.1 | 145 | 0 | 0 | 0 | ... | ... | ... | ... | 40.2 | 1.3 |

[a] Calculated by convolving the number-magnitude relation of Koo & Kron 1988 with the UVX detection probabilities discussed in §6.1.

Table 4. Observed vs. Predicted Numbers of VRX and BRX Quasars

| Selection Method | Magnitude Bin Limits | Number of Candidates | Spectroscopic Identifications | Confirmed z>3 Quasars | Observational Efficiency | Estimated # of z>3 Quasars[d] | Expected # of z>3 Quasars[e] |
|---|---|---|---|---|---|---|---|
| VRX[a] | 17.5<V<20.5 | 3 | 1 | 1 | 100% | 3 | 2.4 |
| VRX[a] | 20.5<V<22.0 | 3 | 2 | 2 | 100% | 3 | 3.4 |
| BRX[b] | 17.5<R<20.0 | 7 | 4 | 0 | 0% | $2.1\pm0.9$ | 0.60 |
| BRX[b] | 20.0<R<21.0 | 16 | 6 | 3 | 50% | $4.8^{+2.2}_{-1.8}$ | 0.76 |
| BRX[c] | 21.0<R<21.5 | 2 | 0 | 0 | ... | $0.6^{+0.7}_{-0.6}$ | 0.37 |
| BRX[c] | 21.5<R<22.0 | 3 | 0 | 0 | ... | $0.9\pm0.9$ | 0.21 |

[a] Based on U-V/V-R + B-V/V-R color-color diagrams.

[b] Based on B-R/R-I86 color-color diagrams.

[c] Based on B-R/R-I86 + B-R/I75-I86 color-color diagrams.

[d] Observational efficiency of $30^{+17.0}_{-10.8}\%$ assumed for the BRX objects.

[e] Calculated using the luminosity function of Warren, Hewett, & Osmer 1994 and the detection probabilities discussed in §6.2-6.3.